\documentclass[conference]{IEEEtran}
\IEEEoverridecommandlockouts

\usepackage{cite}
\usepackage{amsmath,amssymb,amsfonts}
\usepackage{algorithmic}
\usepackage{graphicx}
\usepackage{textcomp}
\usepackage{xcolor}
\usepackage{float}

\usepackage{stfloats} 
\usepackage{tabularx}
\usepackage{subcaption}
\usepackage{lipsum}

\def\BibTeX{{\rm B\kern-.05em{\sc i\kern-.025em b}\kern-.08em
    T\kern-.1667em\lower.7ex\hbox{E}\kern-.125emX}}
\begin{document}

\title{Keyword spotting using convolutional neural network for speech recognition in Hindi}

\author{\IEEEauthorblockN{Saru Bharti}
\IEEEauthorblockA{\textit{iHub DivyaSampark} \\ Indian Institute of Technology, Roorkee \\ Roorkee, India}
\and
\IEEEauthorblockN{Pushparaj Mani Pathak}
\IEEEauthorblockA{\textit{Mechanical and Industrial Engineering Department} \\ Indian Institute of Technology, Roorkee \\ Roorkee, India}
}
\maketitle

\begin{abstract}
In this study, we investigate the application of keyword
spotting (KWS) in the domain of Hindi speech recognition,
utilizing a dataset comprising 40,000 audio samples. With a
sampling rate of 44kHz and an average duration of 1.9 seconds
per sample, we focus on developing an efficient on-device KWS
system tailored for user-specific queries. Leveraging Convolutional
Neural Networks (CNNs) for classification, we employ
feature engineering techniques to convert raw audio recordings into
Mel Frequency Cepstral Coefficients (MFCCs) as an input for our
network. Our experiments encompass various CNN architectures,
exploring their efficacy in identifying predefined keywords within
the continuous speech stream. Our CNN-based approach achieves
a commendable accuracy rate of 91.79\% through
rigorous evaluation, demonstrating promising performance while
ensuring computational efficiency and user-specific customization
in Hindi speech recognition.
\end{abstract}


\section{Introduction}
In recent years, significant advancements have occurred in
speech recognition technology, notably exemplified by platforms
such as Alexa and Google Assistant. These developments
underscore the pivotal role of speech recognition in
driving recent technological progress.
While numerous models exist for English speech recognition,
the landscape narrows considerably for Indian regional languages.
Specifically, the options are scarce in offline Hindi speech
recognition models, with only a handful
available. However, many of these models tend to be either
larger or less accurate compared to their cloud-based
counterparts. 

In Hindi speech recognition, Pruti et al. \cite{1} developed a system that recognizes isolated words in a speaker-dependent manner. The linear predictive cepstral coefficients are used by the system for feature extraction and the Hidden Markov model (HMM) for recognition. They are designed for two male speakers, focusing on Hindi digits zero to nine. Kumar et al. \cite{2} created a high-performance, speaker-independent system with an accuracy of 94.63\%, trained on five males and three females, with a 30-word vocabulary. Feature extraction was done using Mel Frequency Cepstral Coefficients (MFCC), and speech recognition was implemented using HMM. Mishra et al.\cite{3} proposed a connected digit recognition system in 2011, which is speaker-independent and trained on 40 speakers, with added artificial noise. Feature extraction techniques like Revised Perceptual Linear Prediction (RPLP), Bark Frequency Cepstral Coefficients (BFCC),  Perceptual Linear Prediction (PLP) and  MFCC were used, in addition to HMM for recognition. In 2015, Saksamudre et al.\cite{9} presented a study comparing various individual word recognition systems in Hindi, using K-Nearest Neighbour (KNN) and MFCC, with an 89\% accuracy rate. Mishra et al.\cite{15} provided an overview of Hindi speech recognition in 2010, describing speech production and Hindi phoneme properties. Thakur et al.\cite{10} proposed an Automatic Speech Recognition (ASR) system for Hindi speech with varying Indian accents, implementing feature extraction using MFCCs and classification using HMM in 2013. Saini et al.\cite{11} developed a Hindi ASR system using the Hidden Markov Model ToolKit (HTK) in 2013, which recognizes isolated words from a 113-word vocabulary trained on data from nine speakers. In 2004, Kumar et al.\cite{13} proposed a continuous speech recognition system with a large vocabulary for Hindi. In 2013, Sinha et al.\cite{4} proposed a continuous density HMM for context-dependent Hindi speech recognition, comparing vocabulary sizes in databases and feature extraction methods. Rabiner \cite{7} discussed the theory of HMMs in speech recognition. 

The Mozilla Common Voice Hindi dataset, referenced as \cite{p11}, consists of 17 hours of Hindi speech audio alongside their transcriptions, contributed by volunteers who read aloud specified phrases, as noted in \cite{p12}. The University of California Los Angeles (ULCA)-ASR-data-corpus comprises labelled Hindi audio recordings of 2398.76 hours and an additional 2432.92 hours of unlabeled data. This collection includes a variety of speeches from platforms like DD Vigyan Prasar and various Hindi news channels.

The Open Speech and Language Resources (OpenSLR) platform has released Hindi audio datasets specifically for the MUltilingual Code Switching ASR (MUCS) 2021 challenge and also for the 1111 Hours Hindi ASR Challenge \cite{p14} \cite{p15}. The dataset for the MUCS 2021 challenge includes audio files of roughly 100 hours  derived from Hindi narratives, complete with transcripts. 

In \cite{pp}, Bansal et al. developed a transliterated lexicon by converting Hindi vocabulary into English equivalents using Carnegie Mellon University's (CMU) web-based dictionary \cite{pp17} hereafter encoding them into their ARPABET representations (phonetic transcription codes established by Advanced Research Projects Agency, ARPA)\cite{pp18}. The existing open-source datasets have been taken mostly from YouTube and other audio sources like audiobooks and tutorials. 
There is a dearth of open-source Hindi datasets with proper labelling; our project required digits and specific spoken in Hindi with proper labelling.

Nayyer et al. \cite{cnn_mfc} propose a CNN-based KWS system that uses MFCC as preprocessing. They use heavier resnet models and are using the open-source Google Speech Commands dataset to demonstrate the performance.
Lopez et al. \cite{lolo} has published a comprehensive analysis of various applications, datasets and model architectures for KWS.
Diwan et al. \cite{hin1} introduce a new dataset where they provide 600 hours of transcribed speech for 6 Indian languages. They also provide baseline results for multilingual ASR and code-switching ASR using hybrid DNN-HMM models validated on blind test sets.

OpenAI's whisper model also supports Hindi speech recognition. However, the model size is extremely large and has a long processing time, especially on low-power devices using only a CPU for inference. Also, its accuracy was very low while testing on specific word detection. This made it a sub-optimal model for the application.
To circumvent these issues, we developed our own dataset and trained a model with high accuracy that was robust to background noise.

In this paper, section II briefly describes the dataset, its sample distribution across various classes and the audio file details. Section III explains the model architecture that we have adapted, with the details of its layers and the prepossessing that has been done on the data. Section IV deals with the experimental evaluation, such as how the dataset has been expanded based on augmentation techniques and how the model was trained on our ideology. Section V explains the results, and at last, section VI delves into how this study is useful and how it can be helpful in further research.  

\section{Dataset preparation}
 
The dataset was built from scratch. Recorded individuals with different demographics, genders, native languages, and accents. Each sample size is 1.9
seconds and recorded on the mic with a frequency of 44kHz. Various parameters used for dataset preparation are mentioned in Table 1.

The dataset also includes a diverse negative class covering various indoor and outdoor noises. The negative class has been added to identify when nothing is being spoken. In total, we utilized over 40,000 samples spread across 21 classes.
\\Numbers: 0, 1, 2, 3, 4, 5, 6, 7, 8, 9, 10, 11, 12, 13, 14, 15
\\Words: \textit{ha, nhi, sambandh, vibhag}
\\Misc.: negative class

\begin{table}[htbp]
\begin{center}
\caption{Audio Parameters}
  \scalebox{1.3}{
  \begin{tabular}{||c c c ||} 

 \hline
S. no.  & Parameter & Value   \\ [2.0ex] 
 \hline\hline
 1 & Channel type & mono   \\ 
 \hline
 2 & Input file format & .wav  \\
 \hline
 3 & Bit rate & 704 kbps  \\
 \hline
 4 & Sampling rate & 44k \\
 \hline
 5 & Window size & 23 ms \\ 
 \hline
 6 & Framing periodicity & 10 ms \\
 \hline
 7 & Window used & Hann \\
 \hline
 8 & Number of MFCC coefficients & 13 \\
 \hline
 9 & Energy normalization & True \\
 \hline

\end{tabular}}
\end{center}
\end{table}

\section{Model Architecture}
\subsection{Layers}

In this study, we introduce a robust neural network architecture
designed for audio classification tasks, which leverage
Mel-frequency cepstral coefficients (MFCCs) as input features.
The network employs a series of convolutional layers with
increasing filter sizes, in addition to Rectified Linear Unit (ReLU) activations, batch normalization, max pooling to capture robust features from the MFCCs and dropout for regularization, as shown in Fig. 1. 

\begin{figure*}[hbt!]
    \centering
    \setkeys{Gin}{width=0.75\linewidth}
    \begin{subfigure}[t]{0.75\textwidth}
        \centering
        \includegraphics[width=\textwidth]{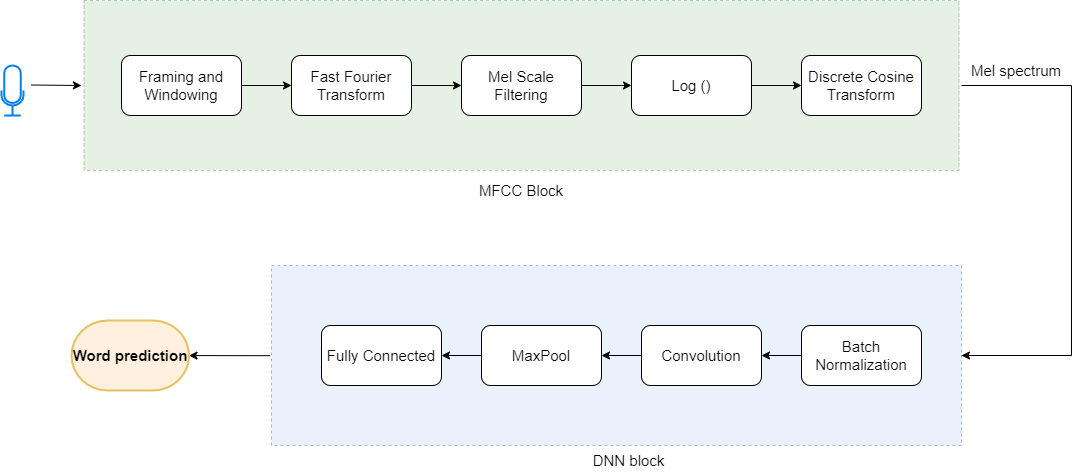}
    \end{subfigure}
   \hfill
   \caption{ System architecture}
\end{figure*}

\begin{figure*}[hbt!]
    \centering
    \begin{subfigure}[t]{0.98\textwidth}
        \centering
        \includegraphics[width=\textwidth]{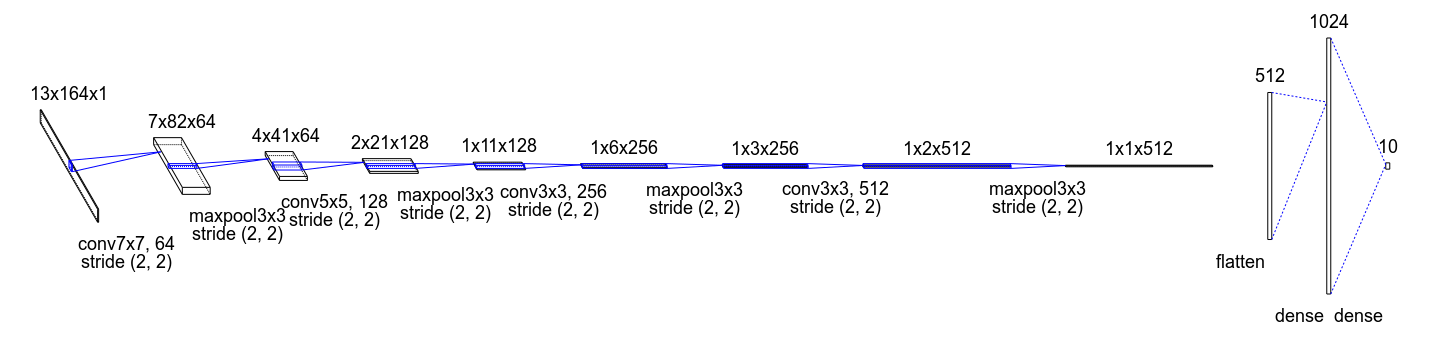}
    \end{subfigure}
   \hfill
   \caption{ Neural network representation}
\end{figure*}

The convolutional layers incrementally increase the receptive field and abstract the spectral features at various levels.
Following the convolutional stack, the architecture includes
two dense layers, consolidating the learned
features into a more abstract representation and facilitating the
classification task. The final layer employs a softmax activation
function to distribute the probability across 21 distinct classes,
ensuring a multi-class classification. This architecture is optimized
for high-dimensional audio data, allowing for precise
categorization while maintaining computational efficiency.

\subsection{Preprocessing}
In the proposed audio classification neural network, we meticulously design the input preprocessing pipeline to capture the essential features of the audio signal. 
The preprocessing pipeline truncates each audio sample to a precise duration of 1.9 seconds. Given a sampling rate of 44 kHz, this corresponds
to $ 1.9 \times 44,000 = 83,600 $ samples per audio clip. 

The preprocessing pipeline then transforms the preprocessed audio into Mel-frequency cepstral coefficients (MFCCs), representing the sound's short-term power spectrum. This transformation is achieved by applying a Fourier transform to a windowed section of the signal, mapping the powers of the spectrum obtained onto the Mel scale, and then taking the logarithm. The Hann window is used to control leakage and increase the dynamic range.
The Hann window is defined as:
\begin{equation}
    W[n] = \begin{cases}
0.5 -0.5\cos\left(\frac{2 \pi n}{N-1}\right) & \text{for } 0 \leq n \leq N-1 \\
0 & \text{otherwise}
\end{cases}
\end{equation}

After the application of window, Discrete Fourier Transform (DFT) is applied on each frame to convert it to the frequency domain:
\begin{equation}
 X[k] = \sum_{n=0}^{N-1} x[n]e^{ \frac{-i2\pi nk}{N}}
\end{equation}

The mel scale is an auditory scale of pitches perceived by listeners to be equidistant from one another. The formula for converting from frequency $ f $ to Mel scale $ m $ is given by:
\begin{equation}
  m = 2595 \log_{10}\left(1 + \frac{f}{700}\right)
\end{equation}

For our neural network, we compute 13 MFCCs  ($ n_{\text{mfcc}} = 13 $) for each frame derived from the log mel spectogram by applying a discrete cosine transform (DCT).  The
window size is 1024, and the hop length, which is the number of
samples between successive frames, is 512. This results
in an overlap of $ 44000 \times 1.9 / 512 - 1 \approx 162 $ frames for each audio clip. The MFCCs are then used as the input feature set for the neural network.

\section{Experimental Evaluation}

\subsection{Data Augmentation}\label{AA}

In this study, we enhance the robustness of our audio classification
neural network through a comprehensive audio dataset
augmentation strategy. Figure 3 and 4 depicts the raw sample waveform and spectrogram respectively. We introduced seven distinct classes of
noise to the audio files, incorporating a variety of everyday
noises—such as vehicular sounds, human conversations, and
environmental noises like flowing water—into our original
audio files, we simulate the acoustic diversity encountered in
real-world settings. Out of 40,000 samples, around 35,000 samples were generated through augmentation.

\begin{figure*}[hbt!]
    \centering
    \begin{subfigure}[t]{0.2\textwidth}
        \centering
        \includegraphics[width=\textwidth]{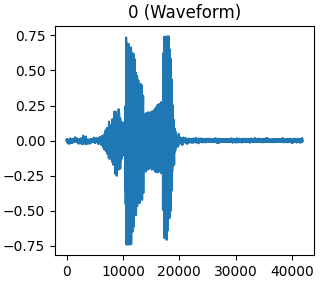}
    \end{subfigure}
    \hfill
    \begin{subfigure}[t]{0.2\textwidth}   
        \centering 
        \includegraphics[width=\textwidth]{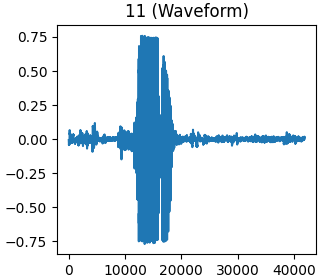}
    \end{subfigure}
    \hfill
    \begin{subfigure}[t]{0.2\textwidth}   
        \centering 
        \includegraphics[width=\textwidth]{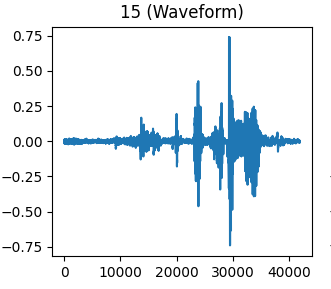}
    \end{subfigure}
    \hfill
    \begin{subfigure}[t]{0.2\textwidth}   
        \centering 
        \includegraphics[width=\textwidth]{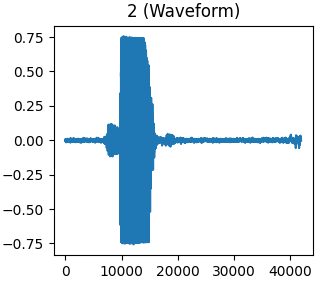}
    \end{subfigure}
    \hfill
    \caption{Raw audio samples waveform}
\end{figure*}

\begin{figure*}[hbtp]
    \centering
    \begin{subfigure}[t]{0.23\textwidth}  
        \centering 
        \includegraphics[width=\textwidth]{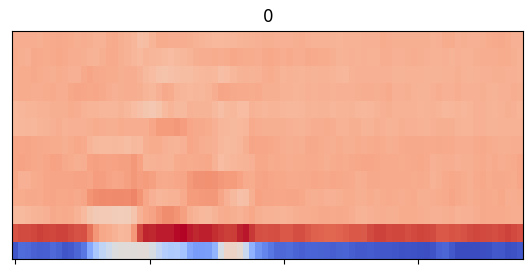}
    \end{subfigure} 
\hfill
\begin{subfigure}[t]{0.23\textwidth}           \centering 
        \includegraphics[width=\textwidth]{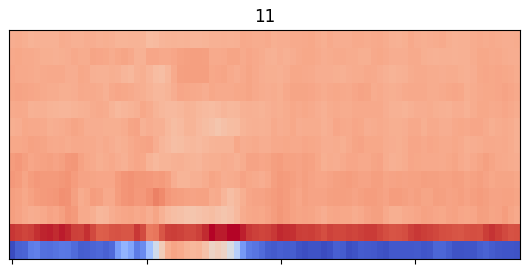}
    \end{subfigure}
    \hfill
    \begin{subfigure}[t]{0.23\textwidth}   
        \centering 
        \includegraphics[width=\textwidth]{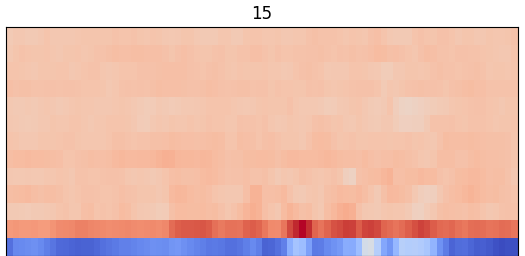}
    \end{subfigure}
    \hfill
    \begin{subfigure}[t]{0.23\textwidth}   
        \centering 
        \includegraphics[width=\textwidth]{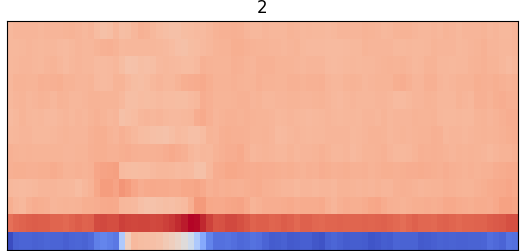}
    \end{subfigure}
    \hfill
    \caption{Raw audio spectrogram}
\end{figure*}

\begin{figure*}[hbt!]
    \centering
    \begin{subfigure}[t]{0.2\textwidth}
        \centering
        \includegraphics[width=\textwidth]{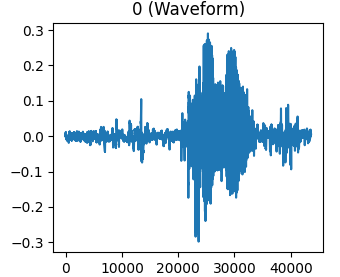}
    \end{subfigure} 
    \hfill
    \begin{subfigure}[t]{0.2\textwidth}   
        \centering 
        \includegraphics[width=\textwidth]{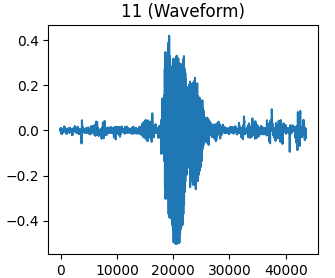}
    \end{subfigure}
    \hfill
    \begin{subfigure}[t]{0.2\textwidth}   
        \centering 
        \includegraphics[width=\textwidth]{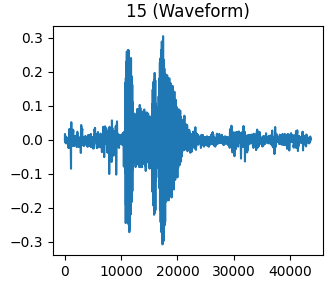}
    \end{subfigure}
    \hfill
    \begin{subfigure}[t]{0.2\textwidth}   
        \centering 
        \includegraphics[width=\textwidth]{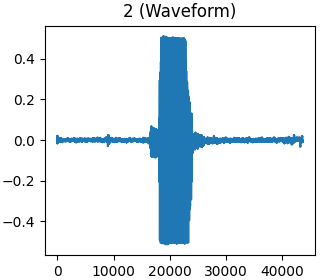}
    \end{subfigure}
    \hfill
    \caption{ Audio samples waveform after augmentation}
\end{figure*}

\begin{figure*}[hbt!]
    \centering
    \begin{subfigure}[t]{0.23\textwidth}  
        \centering 
        \includegraphics[width=\textwidth]{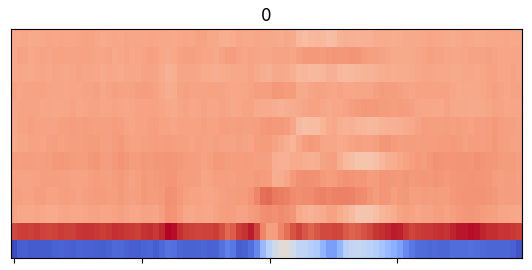}
    \end{subfigure}
\hfill
\begin{subfigure}[t]{0.23\textwidth}           \centering 
        \includegraphics[width=\textwidth]{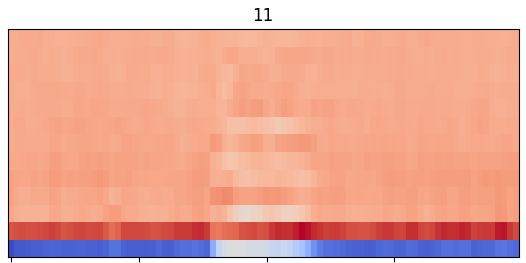}
    \end{subfigure}
    \hfill
    \begin{subfigure}[t]{0.23\textwidth}   
        \centering 
        \includegraphics[width=\textwidth]{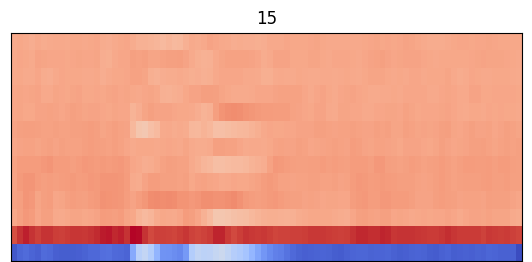}
    \end{subfigure}
    \hfill
    \begin{subfigure}[t]{0.23\textwidth}   
        \centering 
        \includegraphics[width=\textwidth]{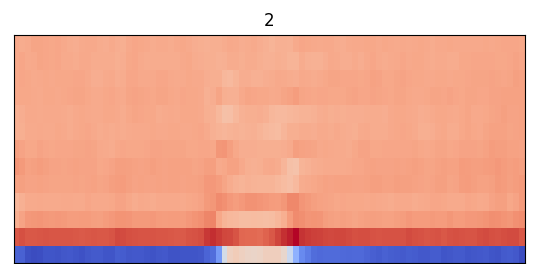}
    \end{subfigure}
    \hfill
    \caption{ Audio spectrogram after augmentation}
\end{figure*}

This augmentation process involves the
careful superimposition of noise samples onto the original audio, followed by volume normalization to maintain consistent audio levels. Additionally, we employ time-shift augmentation
and vary the beginning of the input word and also cover
partially recorded audio to introduce further variability. Through Figures 5 and 6, time shift and addition of noise in the signals due to augmentation can be observed.

These
techniques collectively contribute to the model's improved
performance, elevating its accuracy on real test samples from
60\% to an impressive 91.79\%. The significant enhancement
underscores the efficacy of our dataset augmentation approach
in preparing the neural network to handle the unpredictable
nature of real-world audio inputs.

\subsection{ Methodology}
A supervised learning approach was used to train the neural network, which used cross-entropy as the loss function. Adam optimizer, which is known for its efficiency in managing sparse gradients and adaptive learning rate capabilities, was used for optimization.

Throughout the training, the dataset was divided into mini-batches of 64 samples to strike a balance between computational efficiency and generalization ability.  The learning rate was dynamically adjusted during training for the model, to achieve optimal convergence. Furthermore, an 80:20 split was maintained between training and validation data to ensure a sufficient number of training samples while minimizing the risk of over-fitting. 

This approach resulted in the model achieving 95\% validation accuracy and 91.79\% test accuracy, as shown in the confusion matrix in Figure 7, thus demonstrating its effectiveness in real-world scenarios.

\begin{figure}[htbp]
    \begin{subfigure}[t]{0.52\textwidth}
        \centering
        \includegraphics[width=\textwidth]{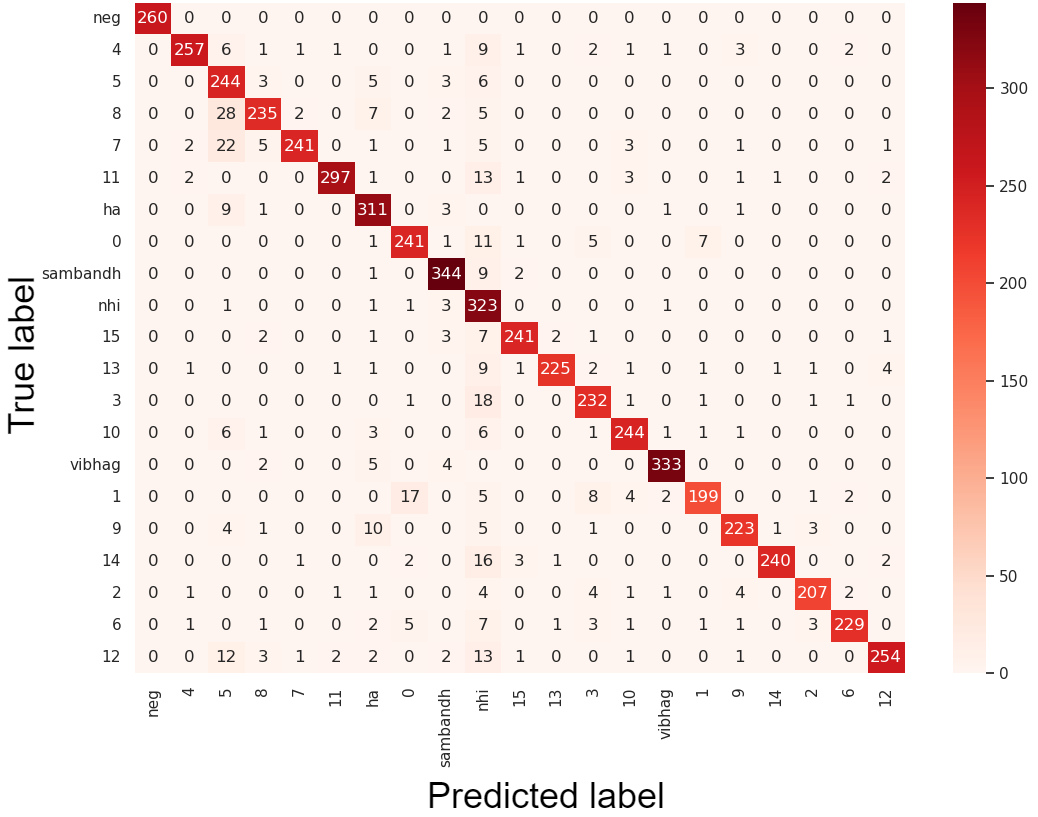} 
    \end{subfigure}   
       \caption{Confusion matrix for model inference }
\end{figure}

\begin{figure}[htbp]
    \begin{subfigure}[t]{0.50\textwidth}
        \centering
        \includegraphics[width=\textwidth]{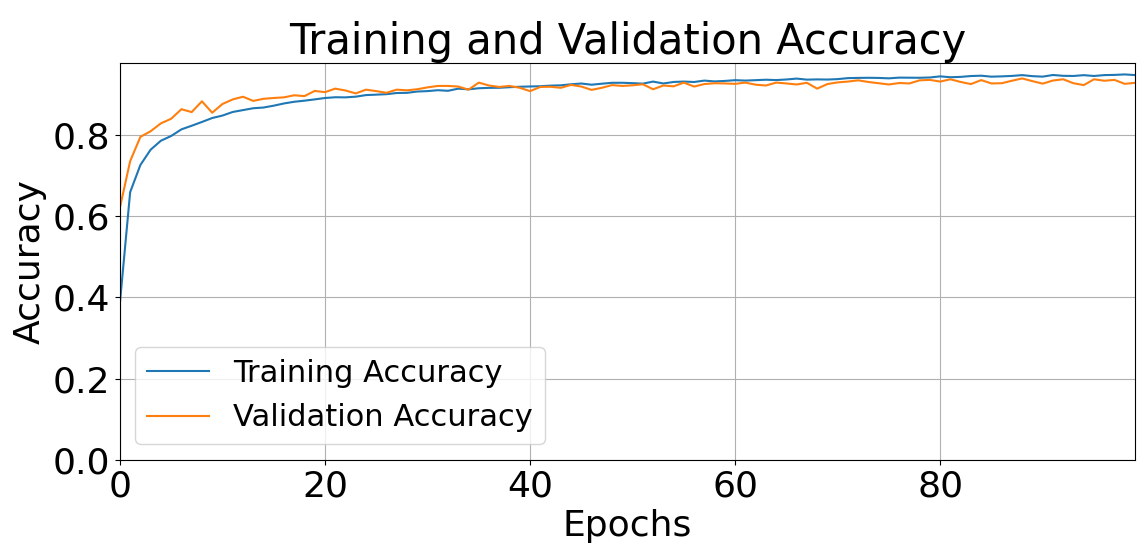} 
    \end{subfigure}   
       \caption{Training and validation accuracy }
\end{figure}

\begin{figure}[htbp]
    \begin{subfigure}[t]{0.50\textwidth}
        \centering
        \includegraphics[width=\textwidth]{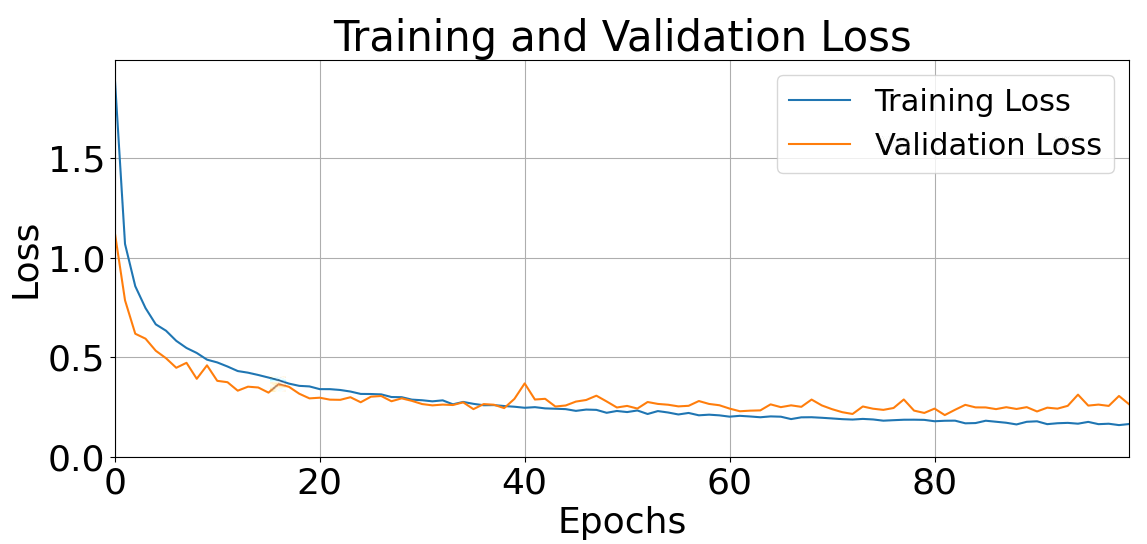} 
    \end{subfigure}   
       \caption{Training and validation loss }
\end{figure}

\section{Results}
Ten samples per person were collected across 21-word classes from a separate group of individuals, thus ensuring the testing of the model in real-life scenarios. These samples were not part of the training process. The final performance of the model on this set was 91.79\%. Along with this, in case of wrong input or no input from the user, the model was able to correctly identify the negative class.

Training and validation graphs for the same are shown in Figures 8 and 9.

\section{Conclusion}
In summary, our research introduces a novel Hindi
keyword dataset alongside a sophisticated neural network architecture  designed to effectively utilize Mel-frequency cepstral coefficients (MFCCs) for robust audio classification.

The model has been carefully tested across various real- world conditions, from scenarios with background noise from human activity to a fan running at maximum speed. Remarkably, the model exhibits an impressive accuracy rate of 91.79\% under such conditions.

Furthermore, the proposed model exhibits remarkable discriminative
capabilities across 21 distinct classes, highlighting its superior precision in categorizing audio signals.

This model can be expanded in the future with a larger dataset of other numbers and words. Also, we can use upcoming neural processing devices for more efficient and faster inference of neural networks in real-time.

\section{Acknowledgement}
We gratefully acknowledge the time and support given by the Robotics and Control lab members for creating the dataset.

\end{document}